\documentclass{article}
\usepackage{spconf,amsmath,graphicx,amssymb,multirow,booktabs,lineno,pgfplots}
\usepackage[noend]{algorithmic}
\usepackage{algorithm}

\title{t-SOT FNT: streaming multi-talker ASR with text-only domain adaptation capability}
\name{Jian Wu, Naoyuki Kanda, Takuya Yoshioka, Rui Zhao, Zhuo Chen, Jinyu Li}
\address{Microsoft, One Microsoft Way, Redmond, WA, USA}
\begin{document}
\ninept
\maketitle
\begin{abstract}
Token-level serialized output training (t-SOT) was recently proposed to address the challenge of streaming multi-talker automatic speech recognition (ASR). T-SOT effectively handles overlapped speech by representing multi-talker transcriptions as a single token stream with $\langle \text{cc}\rangle$ symbols interspersed. However, the use of a naive neural transducer architecture significantly constrained its applicability for text-only adaptation. To overcome this limitation, we propose a novel t-SOT model structure that incorporates the idea of factorized neural transducers (FNT). The proposed method separates a language model (LM) from the transducer's predictor and handles the unnatural token order resulting from the use of $\langle \text{cc}\rangle$ symbols in t-SOT. We achieve this by maintaining multiple hidden states and introducing special handling of the $\langle \text{cc}\rangle$ tokens within the LM. The proposed t-SOT FNT model achieves comparable performance to the original t-SOT model while retaining the ability to reduce word error rate (WER) on both single and multi-talker datasets through text-only adaptation.

\end{abstract}
\section{Introduction}
Multi-talker speech recognition continues to pose a significant challenge because of the serious performance drop on overlapping speech for a conventional single-talker ASR model \cite{barker2018fifth}.
The impact of overlapping speech is significant even with a small ratio of speech overlaps \cite{chen2020continuous,raj2021integration}. 
With the advances of the end-to-end (E2E) ASR technique \cite{zhang2020transformer,chen2021developing,gulati2020conformer,li2022recent}, several efforts have been made for development of the E2E streaming multi-talker ASR model, e.g., SURT \cite{lu2021streaming,raj2022continuous}, MS-RNN-T \cite{sklyar2021streaming}, MT-RNN-T \cite{sklyar2022multi} and t-SOT \cite{kanda2022streaming}. Compared with the earlier modular systems \cite{wu19d_interspeech,yoshioka2018multi,kanda2019guided}, where a speech separation module was employed to address overlapping speech, those E2E ASR solutions transcribe the multi-speaker audio directly, which makes the model simple for both optimization and deployment and potentially brings the better performance.

Among the recent studies, t-SOT models with a Transformer \cite{vaswani2017attention} transducer \cite{graves2012sequence} structure achieved the state-of-the-art (SOTA) recognition results in multi-talker recognition on several datasets including LibriCSS \cite{chen2020continuous,kanda2022streaming} and AMI \cite{carletta2005ami,kanda2023vararray}. Unlike the prior models \cite{lu2021streaming,raj2022continuous,sklyar2022multi} that used two branches for the simultaneous transcriptions of the overlapping speech, t-SOT has only a single output branch to generate the token sequence from multiple speakers.
To distinguish the token streams from different speakers, a special ``channel change'' token, $\langle \text{cc}\rangle$, is inserted when adjacent tokens belong to different speakers.
This simple framework achieved a streaming multi-talker ASR with a simpler model architecture and lower decoding cost. Meanwhile, it also achieved comparable  WER on non-overlapping speech with a single-speaker ASR model, evading the performance degradation witnessed in other ASR \cite{raj2022continuous} and earlier cascaded approaches \cite{chen2021continuous,wu2021investigation}.

While t-SOT models achieved promising results, it still faces a challenge when we adapt it to a specific domain by using only text data.
The challenge stems from its E2E architecture as well as the introduction of special token $\langle \text{cc}\rangle$ to distinguish overlapping speakers.
Firstly, it is known that the E2E ASR model is difficult to adapt by only using only text data, and many researches have been conducted.
Among them, 
the most popular approach  is the language model (LM) fusion \cite{kannan2018analysis,mcdermott2019density,variani2020hybrid,meng2021internal} that incorporates an external LM score on the target domain. However, the performance of these approaches are sensitive to the weight tuning on a development set. 
Recently, factorized neural transducer (FNT) \cite{chen2022factorized,zhao2023fast} was proposed to address this issue. In the FNT model, the prediction network for the regular vocabulary tokens acts as a standard LM. Therefore, various LM adaptation techniques could be applied with the text-only corpus.
However, it is not straightforward to integrate neither the LM fusion nor FNT with t-SOT.
Specifically, a t-SOT model generates tokens spoken by all the speakers in a chronological order together with $\langle \text{cc}\rangle$ token.
The intermingled word sequences from various speakers, along with the $\langle \text{cc}\rangle$ token, disrupt the inherent order of natural language. Consequently, it leads challenges for the standard LM to handle the decoding sequence of the t-SOT model properly, either through LM fusion or FNT.

In this paper, we proposed a novel factorized neural transducer structure named t-SOT FNT to enable the text-only adaptation while maintaining the advantages of the t-SOT based multi-talker ASR. Our changes include two aspects. Firstly,
we change the joint network of FNT to output not only the probability of $\langle \text{blank}\rangle$ token but also that of $\langle \text{cc}\rangle$ token. 
Secondly, $N$ hidden states are maintained within the vocabulary predictor for switching among $N$ concurrent speakers, where $N = 2$ in this work. When $\langle \text{cc}\rangle$ token is emitted, vocabulary predictor will turn to use the other hidden state for the future inference and give all-zero output for the current step. Our experiments show that, compared to a naive t-SOT model, the proposed t-SOT FNT model can achieve comparable performance on meeting conversation data, while achieving better WER on general single-talker ASR set by leveraging a better initialization of the vocabulary predictor network. Moreover, we observe that further WER reduction on both single and multi-talker datasets can be achieved through the text-only adaptation on vocabulary prediction network.

\section{Related Works}

\begin{figure}[!tbp]
\centering
\vspace{-4mm}
\includegraphics[width=0.35 \textwidth]{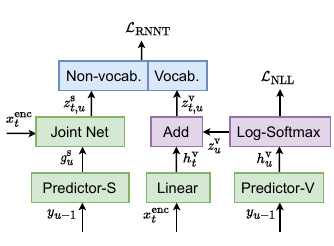}
\vspace{-4mm}
\caption{Illustration of the factorized neural transducer structure. $x_t^{\text{enc}}$ is the $t$-th frame feature from encoder and $y_{u - 1}$ is the predicted token of the previous step. Only green blocks have the trainable parameters.}
\end{figure}

\subsection{Token-level Serialized Output Training (t-SOT)}

The basic idea of the t-SOT is to generate the tokens of all the speakers in a single sequence with a streaming fashion, according to the order of the token emission time in the audio. To achieve this, the serialized transcriptions are generated as the supervision of the t-SOT model. In the case of up to 2 concurrent speakers, a special token $\langle \text{cc}\rangle$ is inserted between the consecutive two tokens once they are spoken by different speakers, indicating the change of the virtual channel, i.e. active speaker. During inference, a post-processing step can be employed to form the transcriptions of two channels from the t-SOT decoding result by switching the output channel index when $\langle \text{cc}\rangle$ is emitted. For example, a decoding sequence of ``\textit{hello how are $\langle \text{cc}\rangle$ i am $\langle \text{cc}\rangle$ you $\langle \text{cc}\rangle$ fine thank $\langle \text{cc}\rangle$ good $\langle \text{cc}\rangle$ you}" can be reformatted to ``\textit{hello how  are you good}" and ``\textit{I am fine thank you}" for the further processing. 
As the t-SOT only changes the supervised label of the ASR model, with the network structure, training loss function consistent as in conventional ASR. Various improvements from one speaker ASR can be easily integrated into t-SOT framework for further performance improvement, such as better network architecture, training scheduler etc.
Meanwhile, we can generalize the t-SOT to support the mixed audio of up to more than 2 concurrent speakers by adding more special ``channel change" tokens. However, given  the overlaps of two speakers are the most common case in the real environment, we continue to  prioritize our efforts on the scenario with 
up to 2 concurrent speakers.

\subsection{Factorized Neural Transducer (FNT)}
FNT decomposes the posterior prediction of the output token set $\mathcal{O}$ into two parts, i.e., $\mathcal{O} = \{\mathcal{V} \cup \mathcal{S}\}$, where $\mathcal{V}$ and $\mathcal{S}$ refer to regular vocabulary tokens and special non-vocabulary token $\langle \text{blank} \rangle$, respectively. As shown in Figure.1, given the $t$-th frame output from the acoustic encoder $x_t^{\text{enc}}$ and the predicted token $y_{u - 1}$ in the previous step, the prediction of the $\langle \text{blank} \rangle$ token follows the standard transducer framework by estimating $z_{t,u}^s$ as
\begin{equation}
z_{t,u}^s = \mathtt{Joint}(x_t^{\text{enc}}, g_u^s),
\end{equation}
where $g_u^s$ is the output of the special prediction network with a input of $y_{u - 1}$. The regular vocabulary prediction $z_{t,u}^v$ is 
computed from $x_t^{\text{enc}}$ 
and the vocabulary predictor output $h_u^v$ as followings. 
\begin{equation}
\begin{aligned}
h_t^v & = \mathtt{Linear}(x_t^{\text{enc}}), \\
z_u^v & = \mathtt{LogSoftmax}(h_u^v), \\
z_{t,u}^v & = h_t^v + z_u^v.
\end{aligned}
\end{equation}
Finally, $z_{t,u}^s$ and $z_{t,u}^v$ are concatenated to form the distribution over $\mathcal{O}$ for the training with the transducer loss $\mathcal{L}_\text{RNNT}$. Besides, a negative log likelihood (NLL) loss is also applied on $z_u^v$ to enforce the vocabulary predictor to act as a standalone LM \footnote{It's equivalent to the way adopted in previous FNT work \cite{chen2022factorized,zhao2023fast} that applied cross entropy (CE) loss on vocabulary predictor's output $h_u^v$.}. As a result, the final objective function for FNT is defined as
\begin{equation}
\mathcal{L}_\text{FNT} = \mathcal{L}_\text{RNNT} + \lambda \cdot \mathcal{L}_\text{NLL},
\end{equation}
where $\lambda$ is a hyper-parameter to control the weight of the LM loss.  During the text only adaptation, the vocabulary predictor is adapted to a target domain based on the available text data. As prior work \cite{zhao2023fast} has shown that adding Kullback-Leibler (KL) divergence loss between the outputs of the adapted vocabulary predictor and original ones can help to avoid the performance degradation on the general domain, the final objective function for text-only adaptation of the vocabulary predictor is
\begin{equation}
\mathcal{L}_\text{adapt} = \mathcal{L}_\text{NLL} + \omega \cdot \mathcal{L}_\text{KLD},
\end{equation}
where $\omega$ is the weight of the KL divergence loss.

\section{t-SOT FNT}

\begin{figure}[!tbp]
\centering
\vspace{-4mm}
\includegraphics[width=0.48 \textwidth]{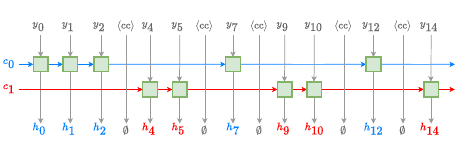}
\caption{The behavior of the vocabulary predictor in proposed integrated t-SOT FNT. $\{y_0, \cdots, y_{14} \}$ are input tokens and $c_0,c_1$ are two hidden states. The green block indicates one processing step of the vocabulary predictor. $\langle \text{cc} \rangle$ serves as the “switch” of two hidden states used for the
vocabulary predictor and once it's emitted, the output of the vocabulary predictor is assigned as all-zero vector.}
\end{figure}

A naive way to incorporate t-SOT framework into FNT architecture is
training FNT by using t-SOT based transcription by treating
$\langle \text{cc} \rangle$ token as a member of regular vocabulary tokens $\mathcal V$.
We refer this combination as ``{\bf naive t-SOT FNT}''.
The naive t-SOT FNT
can achieve multi-talker ASR while keep using the original FNT architecture.
However,
there are two obstacles in this naive combination. Firstly, t-SOT transcription includes an additional channel switching token, i.e $\langle \text{cc} \rangle$, which does not contain semantic meaning and will potentially introduce disruption to the LM.
Secondly, as the serialized outputs in t-SOT mixes transcription from multiple speakers, which breaks the inherent order of natural language and conflicts with the standard LM optimized for single speaker data. As such, the naive combination of FNT and t-SOT will results in inferior performance
and difficulty to incorporate external LM into the vocabulary predictor.

To address the above limitations, we propose a new variant of FNT, 
named {\bf ``integrated t-SOT FNT''}.
In the integrated t-SOT FNT framework, we treat $\langle \text{cc} \rangle$ as a member of special non-vocabulary tokens $\mathcal{S}$, and predict it by the joint network instead of the vocabulary predictor.
In addition, to enable effective integration of external LM into
the vocabulary predictor,
we 
introduce the special procedure to
handle multiple LM states in the vocabulary predictor.

The behavior of the proposed 
vocabulary predictor is exemplified in Figure 2. 
Unlike the conventional FNT as well as the naive t-SOT FNT, the integrated t-SOT FNT maintains multiple hidden states within the vocabulary predictor. 
The number of states to maintain is equal to the maximum number of concurrently active speaker, which is pre-defined to be 2 in this work. 
The $\langle \text{cc} \rangle$ from the decoded sequence serves as the ``switch'' of two hidden states used for the vocabulary predictor,
which enables  
each hidden state to capture the semantic transition of tokens from single speaker, as in the standard LM.
In this way,
an external LM can be smoothly integrated into the t-SOT framework. 


\begin{algorithm}[t]
\caption{Inference process of the vocabulary predictor in t-SOT FNT with up to two concurrent speakers}
\footnotesize
\begin{algorithmic}[1]
\STATE{\textbf{Input}: $\mathcal{Y} = \{y_0, \cdots, y_{U - 1}\} $} 
\STATE{\textbf{Initialize}: $c_0 = \mathbf{\emptyset}, c_1 = \mathbf{\emptyset}, \mathcal{H} = \{\}, j = 0$}
\FOR{$u \;\mathrm{in}\; \{0, \cdots, U - 1\}$}
        \IF{$u = 0$}
        \STATE{$h_u^v, c_0 = \mathtt{PredictorV}(y_u, c_0)$}
        \STATE{$c_1 = c_0$}
        \ELSE
        \IF{$y_u = \langle \text{cc} \rangle$}
        \STATE{$j = 1 - j$}
        \STATE{$h_u^v = \emptyset$} 
        \ELSE
        \STATE{$h_u^v, c_j = \mathtt{PredictorV}(y_u, c_j)$}
        \ENDIF
        \ENDIF
    \STATE{Append $h_u^v \to \mathcal{H}$}
\ENDFOR
\STATE{\textbf{return} $\mathcal{H}$}
\end{algorithmic}
\end{algorithm}

 
More concretely, the inference process of the vocabulary predictor on the intermingled token sequence $\mathcal{Y} = \{y_0, \cdots, y_{U - 1}\} $ is defined in the Algorithm 1, where the current hidden state $c_j$ alternates between $c_0$ and $c_1$.
In transducer, as $\mathcal{Y}$ always starts with the $\langle \text{blank} \rangle$, i.e., $y_0 = \langle \text{blank} \rangle$, we reset the hidden states $c_0, c_1$ at the initial inference step (line 5 and 6). Later when $\langle \text{cc} \rangle$ is emitted, the output of the vocabulary predictor $h_u^v$ is assigned as all-zero vector and the current hidden state $c_j$ for the future inference is switched to $c_{1-j}$ (line 9). For regular vocabulary token, $h_u^v$ is calculated through vocabulary predictor based on the current token $y_u$ and hidden state $c_j$, as shown in line 12. With this design, vocabulary predictor still acts like a standard LM even with the intermingled token sequence that consists of $\langle \text{cc} \rangle$.

In the training stage of the t-SOT FNT, we still use equation (3) as the objective function except that all-zero $h_u^v$ with the $\langle \text{cc} \rangle$ input are masked in the calculation of LM loss, i.e., $\mathcal{L}_\text{NLL}$. 
Thanks to the multi-state design, the text-only adaptation scheme of the integrated t-SOT FNT is same as the standard FNT, which only adapts the vocabulary predictor on the text corpus following the equation (4). 

\section{Experiments Setup}

\subsection{Model Structure}
We used the same encoder structure for all the ASR models in this work, which contained 2 convolution layers and a 18-layer
Conformer \cite{gulati2020conformer} encoder with the chunk-wise streaming mask, resulting in a latency of 160 msec. The attention dimension of the multi-head self-attention (MHSA) layer in each Conformer block was set to 512 with 8 heads and the 2048-dim feed-forward network (FFN)
layer was adopted with the Gaussian error linear unit (GELU).

The prediction network of the single-talker ASR model (referred as ``CT") and t-SOT model (referred as ``t-SOT CT") consisted of a 2-layer 1024 dimensional long short-term memory (LSTM). In both naive and integrated t-SOT FNT models, the non-vocabulary prediction network was a 2-layer 512-dim LSTM and vocabulary predictor was a 2-layer 1536-dim LSTM. We applied a dropout rate of 0.1 to those prediction networks. The dimension of the joint network were all set to 512. The regular vocabulary contained 4003 word pieces thus the output size of single talker model and t-SOT models were 4004 and 4005, respectively.

\subsection{Data and Metric}

We used 30 thousand (K) hours Microsoft in-house data, with the personally identifiable information removed, for the training of the single-talker ASR model. For t-SOT CT and t-SOT FNT model training, we used the combination of the multi-talker simulation data based on the 30K data and real meeting corpus from the training set of AMI \cite{carletta2005ami} and ICSI \cite{janin2003icsi} as well as the Microsoft internal meeting recordings. In the multi-talker simulation using the 30K data, we randomly mixed two utterances on-the-fly with a probability of 67\%. For the rest of 33\%, the original single-talker utterance was used. 
Finally, we used LibriSpeech text data (18 million (M) words),  which was not included in the transcriptions of 30K training data, for the text-only adaptation experiments.

We evaluated our models on several datasets, including the general single-talker ASR test set,
single-distant microphone audio from AMI and ICSI,
and LibriSpeech-style datasets (LibriSpeech \cite{panayotov2015librispeech}, LibriSpeechMix \cite{kanda2020serialized}, LibriCSS \cite{chen2020continuous}).
Our general single-talker ASR test set covers various different application scenarios and consists of a total of 9.9M  words.
It is used to evaluate the single-speaker ASR accuracy before domain adaptation.
On the other hand,
AMI and ICSI were used to evaluate the multi-talker ASR accuracy before domain adaptation. 
We applied a causal logarithmic-loop-based automatic gain control (AGC) on AMI and ICSI to normalize audio volume.
Finally, 
the LibriSpeech-style datasets were used to evaluate the effect of text-only adaptation in both single-talker and multi-talker ASR scenarios. 
We measured WER as an evaluation metric.
For multi-talker test sets, we computed WER based on 
the algorithm proposed in \cite{fiscus2006multiple,kanda2023vararray}.


\begin{table}[!t]
\setlength{\tabcolsep}{2pt}
\footnotesize
\centering
\caption{WER (\%) comparison between t-SOT FNT and other models on general ASR set and AMI and ICSI test set, respectively.}
\begin{tabular}{c|c|c|c|c|c|c|c}
\toprule
\hline
\multirow{2}{*}{\textbf{Model}} & \multicolumn{2}{c|}{\textbf{Seed}} & \multirow{2}{*}{\textbf{General}} & \multicolumn{2}{c|}{\textbf{AMI}} & \multicolumn{2}{c}{\textbf{ICSI}} \\
\cline{2-3}
\cline{5-8}
& \textbf{Enc.} & \textbf{Pred.} & & \textbf{dev} & \textbf{eval} & \textbf{dev} & \textbf{eval} \\
\hline
CT & - & - & 11.9 & 32.9 & 35.9 & 31.7 & 30.9  \\
\hline
t-SOT CT & CT & CT & 12.6 & 21.8 & \textbf{24.6} & 19.7 & 17.4  \\
\hline
Naive t-SOT FNT & t-SOT CT & - & 12.8 &  21.9 & 24.8 & 19.6 & 17.5 \\
\hline
\multirow{4}{*}{Integrated t-SOT FNT} & CT & - & 13.0 & 22.5 & 25.3 & 19.9 & 17.7 \\
\cline{2-8}
& t-SOT CT & - & 12.7 & 21.8 & 24.9 & \textbf{19.2} & 17.2 \\
\cline{2-8}
& CT & LM & 12.8 & 22.6 & 25.6 & 20.1 & 18.3  \\
\cline{2-8}
& t-SOT CT & LM & \textbf{12.2} & \textbf{21.8} & \textbf{24.6} & 19.3 & \textbf{17.1} \\
\hline
\bottomrule
\end{tabular}
\end{table}

\begin{table*}[!t]
\setlength{\tabcolsep}{6pt}
\footnotesize
\centering
\caption{Text-only adaptation results of t-SOT FNT on LibriSpech, LibriSpeechMix and LibriCSS datasets. We used the integrated t-SOT FNT in this experiment, and refer it as t-SOT FNT.}
\begin{tabular}{c|c|c|c|c|c|c|c|c|c|c|c|c|c|c}
\toprule
\hline
\multirow{2}{*}{\textbf{Model}} & \multicolumn{2}{c|}{\textbf{Seed}} & \multirow{2}{*}{\textbf{Adapt}} & \multicolumn{2}{c|}{\textbf{LibriSpeech}} & \multicolumn{2}{c|}{\textbf{LibriSpeechMix}} & \multicolumn{7}{c}{\textbf{LibriCSS}}  \\
\cline{2-3}
\cline{5-6}
\cline{7-8}
\cline{9-15}
& \textbf{Enc.} & \textbf{Pred.} & & \textbf{clean} & \textbf{other} & \textbf{dev-2spk} &  \textbf{test-2spk} & \textbf{0L} & \textbf{0S} & \textbf{OV10} & \textbf{OV20} & \textbf{OV30} & \textbf{OV40} & \textbf{Avg.} \\
\hline
CT  & - & - & $\times$ & 5.9 & 11.9 & - & - & 8.8 & 11.7 & 19.3 & 26.3 & 33.2 & 38.6 & 23.0 \\
\hline
t-SOT CT &  CT & CT & $\times$ & 5.9 & 12.1 & 10.9 & 11.0 & 8.8 & 9.3 & 11.7 & 15.7 & 19.9 & 22.6 & 14.7 \\
\hline
\multirow{4}{*}{t-SOT FNT} & \multirow{4}{*}{t-SOT CT} & \multirow{2}{*}{-} & $\times$ & 6.2 & 12.6 & 11.6 & 11.8 & 9.1 & 10.0 & 12.2 & 15.6 & 20.5 & 23.1 & 15.1 \\
\cline{5-15}
& &  & $\checkmark$ & 5.0 & 10.7 & 10.3 & 10.4 & 7.9 & 8.6 & 11.1 & \textbf{14.4} & 18.9 & 22.1 & 13.8 \\
\cline{3-15}
& & \multirow{2}{*}{LM} & $\times$ & 5.5 & 10.9 & 11.0 & 10.6 & 8.6 & 9.0 & 11.2 & 15.4 & 19.9 & 22.5 & 14.5 \\
\cline{5-15}
& & & $\checkmark$ & \textbf{4.7} & \textbf{10.4} & \textbf{10.1} & \textbf{10.1} & \textbf{7.9} & \textbf{8.2} & \textbf{10.5} & 14.5 & \textbf{18.8} & \textbf{21.8} & \textbf{13.6} \\
\hline
\bottomrule
\end{tabular}
\end{table*}

\subsection{Training and Evaluation}
The 80-dim log mel-filterbank using 25 msec window and 10 msec hop size was extracted as the input feature for ASR models. We applied global mean and variance normalization. All the ASR models were trained on 16 NVIDIA V100 GPUs with AdamW optimizer. For CT model, we performed 500K-step training with a linear decay learning rate scheduler. 50K warm-up steps were used and the peak learning rate was set to $1.5e^{-3}$. t-SOT CT model was trained for 275K steps, using the CT model as the seed. Warm-up steps were removed and the peak learning rate was set to $2e^{-4}$.

To simplify the training scheme of t-SOT FNT, the encoder parameters were initialized from the well-trained CT or t-SOT CT models, and the peak learning rate was set to $4e^{-4}$  and $2e^{-4}$, respectively. Other configurations were kept same as t-SOT CT, including the training steps and batch size.
Following the previous FNT work \cite{zhao2023fast}, in order to utilize more text data, we can also initialized vocabulary predictor of the FNT from a pre-trained LM with same architecture but was trained independently on a much larger
text corpus. In this work, the text corpus for LM training contains 29 Billion words.
When it comes with text-only adaptation, we adapt the vocabulary predictor of the integrated t-SOT FNT model for 10K steps on 4 V100 GPUs with a learning rate decayed from $1e^{-4}$ to 0. $\omega$ was set to 1 to keep the performance on general ASR test set. For ASR decoding, we used a beam size of 16.

\section{Results}

In this section, we first discuss the performance of the t-SOT FNT before adaptation on single and multi-talker test sets in Section \ref{sec:result1}. We then discuss the text-only adaptation results  in Section \ref{sec:result2}.

\subsection{Performance of single and multi-talker ASR}
\label{sec:result1}
The WERs on the general single-talker ASR set, AMI and ICSI are reported in Table 1, where four model structures are listed and compared. 
With the same training configuration, our proposed integrated t-SOT FNT model outperformed the naive t-SOT FNT on ICSI dataset while achieving comparable WERs on AMI and general single-talker ASR test sets. This result illustrates that the proposed multi-state vocabulary predictor of the integrated t-SOT FNT works on par with the vocabulary predictor of the conventional FNT, while the former provides a way to naturally integrating external LM into the
vocabulary predictor.


Among the integrated t-SOT FNT variants, we observed that starting training from t-SOT CT encoder leads better performance than that from CT. This is expected as the former one has been optimized on multi-talker data. On the other hand, initializing the vocabulary predictor with a pre-trained LM on the larger scale text data resulted better WER on general ASR set, but the improvement was limited on AMI and ICSI.  The reason might be that the training corpus of the LM covers the scenarios of the general ASR set but lacks sufficient meeting conversation data.

On general single-talker ASR test set, the best t-SOT FNT model (last row) achieved better result
than t-SOT CT,
closing the WER gap from the single-talker CT from 0.7\% (=12.6\%-11.9\%) to only 0.3\% (=12.2\%-11.9\%). On AMI data, t-SOT FNT achieved similar performance with t-SOT CT while on ICSI, it outperformed t-SOT CT model by a 0.4\% and 0.3\%  on development and evaluation sets, respectively. These results
demonstrated the capability to convert an existing t-SOT CT model to a t-SOT FNT model by keeping the accuracy of the original t-SOT CT model.

\subsection{Results of the text-only adaptation}
\label{sec:result2}
We picked the top two integrated t-SOT FNT models from Table 1 and performed text-only adaptation using text data from LibriSpeech training set. The results were reported in Table 2, where three data sets, LibriSpeech, LibriSpeechMix, and LibriCSS, were evaluated. The performance of the t-SOT FNT were improved after adaptation, not only on single-talker audio, but also on the multi-talker data regardless of if the data is simulated mixture (LibriSpeechMix) or real mixture (LibriCSS). Overall, the t-SOT FNT with LM initialization achieved best performance.
Compared with t-SOT CT, t-SOT FNT brought a relative WER reduction of 8.4\% and 7.5\% on LibriSpeechMix and LibriCSS, respectively. In addition, compared with CT that achieved
5.9\% and 11.9\% on LibriSpeech, t-SOT FNT achieved
significantly better WERs of 4.7\% and 10.4\% by enjoying the text-only adaptation capability.

Among the t-SOT FNT models, we observed that the text-only adaptation closed the gap between the LM initialization on all the three test sets. For example, the relative averaged WER difference on LibriSpeechMix was reduced from 8.3\% to 2.5\%, and that on LibriSpeech was reduced from 14.6\% to 4.0\%. Overall, our results demonstrated that the proposed t-SOT FNT enjoyed 
the advantage of 
the vocabulary predictor where the general single-talker ASR accuracy was improved by utilizing a powerful LM, and the accuracy was further improved by the text-only domain adaptation. 

\section{Conclusions}
In this paper, we proposed the t-SOT FNT model to incorporate the text-only adaption capability into the multi-talker ASR. A set of the hidden states were maintained within the vocabulary predictor to keep track the natural token transition from non-overlapping speakers. Compared with t-SOT CT model, the proposed t-SOT FNT achieved comparable WER on AMI and ICSI data sets and better WER on general single-talker ASR set. The experiments on LibriSpeech-style test set further demonstrated that significant WER reduction can be obtained by text-only domain adaptation on both single-talker and multi-talker audio.

\bibliographystyle{IEEEbib}
\bibliography{main}

\end{document}